# The Objectives of the Radioscience Experiment in Luna-Resource and Luna-Glob Space Projects.

**Gromov, V. D. , Kosov, A. S.** , *IKI RAS, 84/32Profsouznaya, Moscow, Russia. Contact: vgromov@iki.rssi.ru*

**Introduction:**
Measurements of the Doppler shift of a radio signal from a spacecraft, along with interferometric location of angular position of the radio source, achieves nowdays a very high accuracy. It permits to use them not only for orbit determinations, but also for sophisticated research of mass and gravity effects. A last decade demonstrated an impressive progress in these investigations [1-3]

It is a frequently situation when achievement of some high level of accuracy of measurements produces a break-through in science, like it was with plentiful exoplanet detections due to progress in photometric accuracy, or with determination of most fundamental parameters of Universe due to achievement by the Radioscience community of an accuracy $\sim 10^{-6}$ in measurements of intensity of the Cosmic Microwave Background [4-6]. In space Doppler measurements the accuracy is approaching to $10^{-8}$ (better than 0.03 mm/sec).

**Radioscience Measurements in Lunar Missions and Accuracy Limitation Factors:**
*Orbital Doppler Measurements.*
Instrument uncertainties of Doppler measurements are linked with instability of local oscillators and thermal noises of radio receivers. For the ground based stations for space communications, there is an additional source of errors, fluctuations of signal delay in Earth troposphere and ionosphere. Decreasing of the ionospheric errors could be reached by increasing of the carrier frequency, particularly by a transition from X-band to Ka-band. Doppler measurements on orbit of the Moon entirely exclude the Earth environment effects. Therefore the orbiter's Ka-band receiver can measure acceleration with high accuracy. The acceleration variations are related to deviations of Luna's gravitation field. Precise measurements of Lunar gravity was made in GRAIL experiment [2, 3] using two orbiters. The accuracy of the Lander-Orbiter experiment could be better, but a coverage of the Lunar surface is more restricted.

*VLBI Interferometry.*
Very Long Base Interferometry is a traditional tool for precise measurements of a descent/lander probe position since the Vega mission [7, 8]. An extended VLBI network permits to measure the Lander's position with high accuracy in X-band in spite of tropospheric and ionospheric errors.

*Same Beam Interferometry (SBI).*
For SBI experiments, several radio beacons on Moon surface should work simultaneously and be synchronized from a single Earth's reference source. The beacon signals are received by the same antenna beam, so the tropospheric and ionospheric effects in measurements of relative position of the beacons are efficiently suppressed.

**Scientific phenomena, for which investigations the Radioscience is a most fruitful tool:**
*Orbital and rotational movement of the Earth and the Moon.*
A movement of lunar probes is determined by positions of Earth, Moon and other Solar System bodies. Precise measurements of probe velocities and positions gives information for improvement of the reference frames for the Earth and the Moon, for better lunar rotation dynamics, lunar orbit, and lunar ephemeris, incorporating the numerous Newtonian perturbations as well as the much more subtle relativistic phenomena, which are all used for spacecraft missions.

*Mass distribution and possible internal movements in the Moon's interior.*
A movement of orbital probe is sensitive to subsurface mass concentration. A movement of lander probe reflects rotation of the Moon's body, revealing its precession, nutation and wobbling, which are dependent on the moments of inertia of the whole Moon and the structure of its core (inner core and outer core).

*General Relativity (GR) effects.*
Accuracy of Radioscience measurements is enough to reveal GR deviation from Newton theory. More and more precise determination of the GR effects gives the tests for comparison of GR with other theories of gravity.

**Scientific Perspectives for Lunar Radioscience Experiments:**
An accuracy of acceleration measurements in the Lander-Orbiter experiment [9] coud be about 3-10 mGal, two times better than in GRAIL experiment. The area of these measurements is restricted by distance approximately equal to a value of height *H* of an orbit.

The altitude distribution of the gravity potential $V$ of the Moon could be written using an expansion in harmonics:

$$V = \frac{GM}{R+H}\left[-1 + \sum_{n=2}^{m}\left(\frac{R+H}{R}\right)^{-n} P_n(\lambda, \theta)\right],$$

where $G$ is the gravitational constant, $M$ and $R$ are the mass and the radius of the Moon, $\lambda$ and $\theta$ are longitude and latitude on the Lunar surface, $P_n$ are sums of the $n$-th degree Legendre polynomials with coefficients to be defined from measurements for the definition of the nonuniformity of the Lunar gravity field picture. The maximum degree m of the expansion is inversely proportional to the spatial resolution $\Delta L$ of the picture. To a value $\Delta L \sim 10$ km corresponds $n \sim 500$. The equation demonstrates so strong damping with altitude of the high resolution details, that measurements with resolution $\Delta L$ needs an orbit with $H \sim \Delta L$, which reduces a measurement area diameter to $\sim 2H$.

The scientific perspectives of the experiments depend on accuracy of resulting scientific values, obtained after experimental data reduction using a quantitative (digital) model of the physical phenomenon. A model is also necessary for the choice of the experiment configuration (orbit and so on) due to contradictory requirements to some configuration parameters, as was shown above. These scientific models are under development by the Radioscience Experiment Team.